\documentclass[aps,prl,twocolumn,
superscriptaddress,
               amsmath,a4paper,floatfix]{revtex4}

\usepackage{graphicx}
\usepackage{dcolumn}
\usepackage{bm}
\usepackage{dcolumn}
\usepackage{subfigure}

\graphicspath{{../}}

\begin{document}

\title{\textbf{
Spin Hamiltonian, Competing Small Energy Scales and Incommensurate 
Long Range Order in the Highly Frustrated Gd$_3$Ga$_5$O$_{12}$ 
Garnet Antiferromagnet
}}

\author{Taras Yavors'kii}
\affiliation{
Department of Physics and Astronomy, University of Waterloo, Waterloo, Ontario, N2L 3G1, Canada}
\author{Matthew Enjalran}
\affiliation{
Department of Physics, Southern Connecticut State University, New Haven, CT 06515, USA}
\author{Michel J.P. Gingras}
\affiliation{
Department of Physics and Astronomy, University of Waterloo, Waterloo, Ontario, N2L 3G1, Canada}
\affiliation{
Department of Physics and Astronomy, University of Canterbury,
Private Bag 4800, Christchurch, New Zealand}

\date{\today}

\begin{abstract}
Despite the availability of a spin Hamiltonian for the Gd$_3$Ga$_5$O$_{12}$
garnet (GGG) for over twenty five years, there has so far
 been little theoretical 
insight regarding the many unusual low temperature properties of GGG. 
Here we investigate GGG in zero magnetic field 
using mean-field theory. We reproduce the spin liquid-like 
correlations and, most importantly, explain 
the positions of the sharp peaks seen in powder neutron diffraction experiments.
We show that  it is crucial
to treat accurately the long-range nature of the magnetic dipolar 
interactions to allow for a determination
of  the small exchange energy scales involved in the
selection of the experimental ordering wave vector.
Our results show that the incommensurate order in GGG is
classical in nature,
intrinsic to the microscopic spin Hamiltonian and not caused
by weak disorder.
\end{abstract}
\maketitle

The diversity of empirical data 
collected over the past fifteen
years has demonstrated that geometrically frustrated
triangular and tetrahedral
arrangements of antiferromagnetically coupled spins
are highly partial towards the realization of 
exotic correlated phases in magnetic 
materials~\cite{Diep-review,Ramirez-review,Greedan-review}.
The reason for the rich and typically material specific
properties of frustrated magnets is understood.
It stems from their sensitivity to perturbations beyond
the frustrating nearest-neighbor antiferromagnetic (AFM)
exchange which, on its own, leads to a macroscopic number of 
exactly degenerate and competing, hence fragile, 
classical ground states.
In this paper we show, through a careful theoretical analysis
of neutron scattering experiments, that the extensively studied 
Gd$_3$Ga$_5$O$_{12}$ garnet (GGG) is precisely such a system,
though evidence for this fact emerges from a 
perspective on the problem that has heretofore
escaped scrutiny.

GGG displays a gamut of complex and interesting
low temperature magnetic phenomena. 
In zero magnetic field, the behavior of GGG
is uniquely rich.
The nonlinear magnetic susceptibility $\chi_3$
peaks at $T_g\sim 180$ mK~\cite{Schiffer-SG}, 
indicating a spin glass transition~\cite{Mydosh}.
However, muon spin relaxation~\cite{Dunsiger,Marshall}
and M\"ossbauer spectroscopy~\cite{Bonville}
find persistent spin dynamics down to $T\ll T_g$.
Meanwhile,
powder neutron scattering data~\cite{Petrenko-PRL,Petrenko-Physica}
indicate
that GGG is on the verge of developing true
incommensurate
long-range magnetic order with a correlation length 
($\xi \approx 100$ \AA) extending over 8
cubic unit cells below 140 mK.

A Hamiltonian ${\cal H}$ describing GGG, that we shall explicitly define below, 
has long been available~\cite{Wolf}. 
It assumes classical Gd$^{3+}$ spins, is parameterized 
as a sum of empirical exchange contributions up to third nearest-neighbors 
as well as a magnetic dipolar contribution, 
and ignores potentially important quantum fluctuations 
or disorder inherent to GGG~\cite{Schiffer-SG}. 
Previous numerical studies 
based on ${\cal H}$ have been 
unable to provide a quantitative explanation for 
the bulk~\cite{Schiffer-SG}
and dynamical~\cite{Dunsiger,Marshall,Bonville}
properties of GGG or the incommensurate
spin-spin correlations that develop below 
200 mK~\cite{Petrenko-PRL,Petrenko-Physica}.
This could be interpreted as evidence that
exotic mechanisms 
involving either quantum fluctuations or
disorder effects 
are at play in GGG. 
For example, it has been suggested 
that disorder, such as
Gd$^{3+}$ on Ga$^{3+}$ sites, is responsible for nucleating
regions with the spin order 
observed in neutron 
scattering in zero magnetic field~\cite{Petrenko-PRL,Petrenko-Physica}.

In this letter we address the adequacy of the Hamiltonian
${\cal H}$ of 
Ref.~\onlinecite{Wolf} to 
explain the
low temperature properties of GGG. 
It has been demonstrated \cite{Petrenko-PRL,Petrenko-PRB} that 
${\cal H}$ with only a nearest-neighbor exchange interaction
can describe 
liquid-like spin correlations 
\cite{Petrenko-PRL,Petrenko-Physica}.
We show below that ${\cal H}$
can also
describe well the incommensurate spin-spin correlations
and that there is no obvious reason 
to extend ${\cal H}$.
Rather, it is the correct procedure
to treat its many competing small energy scales, in particular the
long range 
dipolar interactions, and the unusually precise
specification of weak
second and third nearest-neighbor exchange couplings,
that are foremost needed, and that have been missed so far. 

In order to correctly parameterize ${\cal H}$ and the resulting
spin correlations we rely on
neutron scattering experiments~\cite{Petrenko-PRL,Petrenko-Physica},
which are the most direct and unambiguous probe of magnetic correlations. 
Our 
basic hypothesis is that
conventional criticality controls the development of the
sharp 
peaks observed in neutron 
experiments~\cite{Petrenko-PRL,Petrenko-Physica}, and that, as in conventional magnets, 
a mean-field 
theory (MFT)
treatment of the pertinent Hamiltonian 
should be
able to capture 
the Gaussian regime in the approach to criticality.
Our finding of a unique ordering wave vector ${\bf q}_{\rm ord}$ shows 
that GGG, unlike the AFM nearest-neighbor Heisenberg pyrochlore, 
is not pathological and is amenable to such a 
soft-mode determination and calculation of 
the neutron scattering intensity profile. We show that GGG is 
very similar to the now well understood 
dipolar spin ice materials~\cite{Gingras-CJP,Melko}, where 
a MFT
that takes into account the
long-range nature of the dipolar interaction~\cite{Gingras-CJP}
(i) identifies the correct ordering wave vector~\cite{Melko},
and (ii) correctly predicts the 
symmetry of the 
scattering 
close to the critical 
temperature.

We examine the spin 
correlations in GGG by calculating 
the powder neutron scattering intensity $I({Q})$ 
and comparing it with experimental data~\cite{Petrenko-PRL}.
The scattering ${\bf {q}}$-vector-dependent
intensity  is  given by
$I({\bf q}) =({F(|{\bf q}|)^2}/N)
 \sum_{ij} \langle {\bf S}_{i}^{\perp}
\cdot{{\bf S}}_{j}^{\perp} \rangle e^{\imath{\bf {q}}\cdot{\bf {r}}_{ij}}$,
obtained for the model Hamiltonian~\cite{Wolf}
${\cal H}={\cal H}_{\rm e}+{\cal H}_{\rm dip}$ with
${\cal H}_{\rm e}= \sum_{i>j} \left\{ 
\sum_{n=1}^3 J_n  \delta_{r_{ij},r_n} {\bf S}_i\cdot{\bf S}_j \right \}$
and 
${\cal H}_{\rm dip}= \sum_{i>j} 
[D(r_1/{r_{ij})^3}] 
\left[ {\bf S}_i\cdot{\bf S}_j - 3 ({\bf S}_i\cdot\hat{r}_{ij}) ({\bf S}_j
\cdot\hat{r}_{ij})\right ]$.
Labels $i$ and $j$ span the $N$ sites of a garnet lattice
(c.f. Fig.~2 in Ref.~\cite{Wolf}), 
$r_{ij}$ and $\hat{r}_{ij}$ are the length and direction of 
the vector separation between them, and
$r_n$ enumerate distances between 
the sites starting from the nearest neighbors.
The size of the GGG conventional cubic unit cell is $a=12.349$ \AA~\cite{Petrenko-PRL}.
The isotropic vectors ${\bf S}_i$ have 
length $\sqrt{S(S+1)}$ to
classically approximate the $S=7/2$ $\rm Gd^{3+}$ spins.
${\bf S}_i^\perp$ are the spin components perpendicular
to ${\bf q}$, $F(|{\bf q}|)$ is the Gd$^{3+}$  
magnetic form-factor \cite{Brown}, and
$\langle \ldots \rangle$ denotes a thermal average.
The spin couplings in ${\cal H}$ include 
dipolar interactions ${\cal H}_{\rm dip}$ 
of strength $D$, 
cut-off at a distance $R_c$, 
and exchange interactions of empirical strength $J_n$, $n\leq3$.
We show
below that $R_c$ must be set to the true long-range limit
$R_c=\infty$ in order 
to obtain physically meaningful results.
Throughout the paper, we keep 
$D=0.0457$~K \cite{Petrenko-PRB}
and the nearest-neighbor exchange 
$J_1=0.107$ K \cite{Wolf}
 fixed.
However, because of the large uncertainties on
$J_2$ and $J_3$ estimated from high-temperature analysis
($-0.015\leq J_2 \leq 0.009$ K, 
$-0.035 \leq J_3 \leq 0.100$ K)~\cite{Wolf},
and the unreliability of the Monte Carlo procedure~\cite{Wolf}
to refine them because of poor statistics 
(see also discussion in Ref.~\cite{Petrenko-Physica}),
we consider 
these
as free parameters
to be adjusted by fitting 
the experimental $I(Q)$
\cite{Petrenko-PRL}. 

\begin{figure}[t]
\includegraphics[angle=-90,width=3.2in]{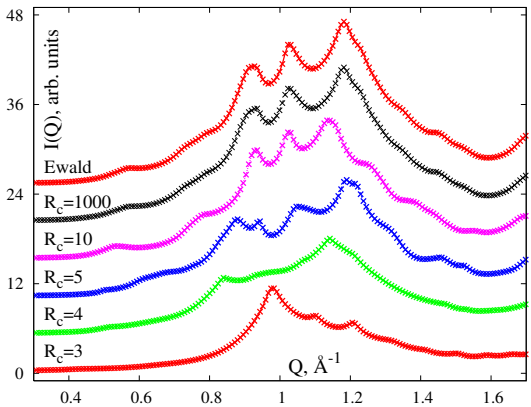}
\caption{
MFT 
$I(Q)$ of GGG with 
$J_2=-0.003$ K, $J_3=0.010$ K 
\cite{Wolf} and 
$\tau=10^{-4}$ 
for 
$R_c=3,4,5,10,1000$.
The $R_c=\infty$ profile is obtained using the Ewald method. 
The profiles are uniformly shifted to emphasize their strong
dependence on $R_c$ for small values, as well as their saturation 
to a well defined limit at large $R_c$.
}
\label{IqRc}
\end{figure}

We use a Gaussian MFT~\cite{MFT}
to obtain the soft modes
and calculate 
$I(Q)$ of the model ${\cal H}$.
Using a self-adaptive ${\bf q}$-grid,
we can calculate $I({\bf q})$ 
(in the thermodynamic limit)
at an arbitrary positive dimensionless temperature 
$\tau=T/T_{\rm c}^{\rm MFT}-1$, where
$T_{\rm c}^{\rm MFT} \approx  2$~K and depends on $J_2$ and $J_3$.
Due to the $\log|\tau|$ growth of the powder MFT 
Bragg peak intensities,
as opposed to the $1/\tau$ growth of the ${\bf q}$-dependent 
intensities, 
$\tau$ 
must be set rather small in order to work with the experimental profiles
in the critical regime (cf. Fig.~\ref{figIq}).
We consider a finite ${\bf q}$-space grid with 
$32^3$ points in the first Brillouin zone,
and construct a three dimensional cubic interpolation scheme
to separately evaluate the numerator and the denominator 
of the MFT $I({\bf q})$ (cf.~Eq.~A37 in Ref.~\onlinecite{MFT}).
For ${\bf q} \approx {\bf q}_{\rm ord}+{\bf G}$, where Bragg peaks
occur 
(${\bf G}$ is a FCC reciprocal lattice vector), 
we back up the interpolation by 
calculating $I({\bf q})$ exactly~\cite{MFT}.
This ensures ${\bf q}$-grid size independent results~\cite{MFT}.
$I(Q)$ is computed by 
spherically averaging $I({\bf q})$ numerically.

We find that ${\cal H}$ with dipolar interactions 
and exchange interactions beyond nearest-neighbor 
possesses a unique ${\bf q}_{\rm ord}$.
Our numerical study suggests that the soft mode
selection in the full ${\cal H}$ is highly
sensitive to the values of $J_2$ and $J_3$. This finding is
in 
general
agreement with results from 
Monte Carlo simulations of GGG~\cite{Petrenko-PRL,Petrenko-Physica}
(see also Ref.~\cite{Wolf} for a comment on the
sensitivity of the GGG specific heat 
on $J_3$).
However, our crucial new observation is that ${\bf q}_{\rm ord}$ is also very
sensitive to 
an ad-hoc choice
of the cut-off distance $R_c$  in ${\cal H}_{\rm dip}$.
We demonstrate this by plotting 
in Fig.~\ref{IqRc} $I(Q)$ for various $R_c$
for $J_2=-0.003$ K and $J_3=0.010$ K 
(values from~\cite{Wolf}).
For $R_c \lesssim {\rm O}(10^2)$, not only is ${\bf q}_{\rm ord}$
a 
non-monotonous
function of $R_c$,
as the positions of the peaks indicate, but so is the
unit cell magnetic form factor, as is evident from the 
significant changes in the overall shape of the diffuse scattering patterns
near criticality ($\tau \ll 1$).
Fig.~\ref{IqRc} demonstrates that ${\cal H}$ with different 
low $R_c$ 
gives completely different spin-spin correlations in the critical regime.
Only for rather large cut-offs does ${\bf q}_{\rm ord}$ become
insensitive to $R_c$, approaching the $R_c=\infty$ limit 
only
when 
$R_c \gtrsim 10^3$.
Figure~\ref{IqRc} suggests that it is essential
to set $R_c$ to its true $R_c=\infty$ limit
to avoid
inducing
spurious ordered phases by
treating the 
lattice sum in ${\cal H}_{\rm dip}$
by a short cut-off method,
as was done 
previously~\cite{Petrenko-PRL,Petrenko-Physica,Petrenko-PRB,Wolf,Schiffer-field}.
We henceforth acknowledge this and implement the Ewald method~\cite{MFT}
to process $H_{\rm dip}$ with $R_c=\infty$.

\begin{figure}[t]
\includegraphics[width=3.2in]{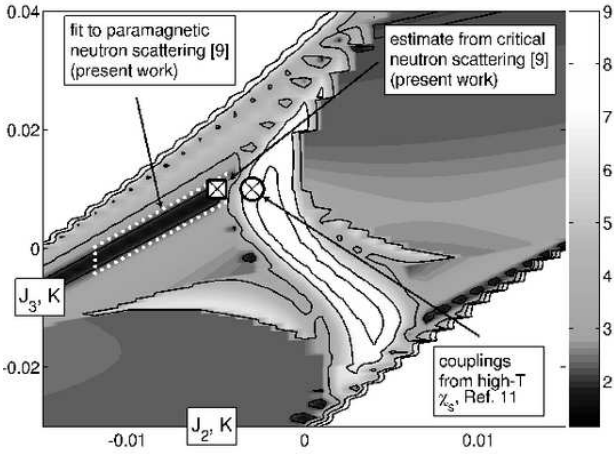}
\caption{
The optimization neutron-scattering penalty 
function, $\mathcal{P}$, 
in the $J_2-J_3$ plane behaves discontinuously and shows minima
in the black domains.
Only for the couplings 
demarcated by a dotted trapezoid
does the theoretical 
scattering profile reconstitute
both the location of the sharp peaks and the overall shape of the broad
and diffuse experimental scattering~\cite{Petrenko-PRL,Petrenko-Physica}.
The circle corresponds to the values of $J_2$ and $J_3$ 
suggested in Ref.~\onlinecite{Wolf}.
The upper-left and bottom-right white regions correspond to
different ${\bf {q}}_{\rm ord}=0$ structures. 
} 
\label{figJ2J3map}
\end{figure}


The scattering profiles in Fig.~\ref{IqRc} 
and the ${\bf q}$ positions of the sharp peaks, 
even in the $R_c=\infty$ limit, are 
incompatible
 with the 
experimental data~\cite{Petrenko-PRL} (cf. Fig.~\ref{figIq}, upper panel).
However, the uncertainty in previously estimated values 
of $J_2$ and $J_3$ \cite{Wolf} allows fine-tuning 
them in order to obtain
a better match with the 
low temperature data~\cite{Petrenko-PRL}.
To proceed,
we are guided by 
the
general 
understanding of disordered  systems 
near the critical boundary between long-range ordered and spin glass 
phases~\cite{Mydosh}.
Specifically, we make the reasonable assumption
that, given the sizeable and saturated
$\xi\sim 100$ \AA\  
correlated regions
below 140 mK ~\cite{Petrenko-PRL}, 
the weak disorder in GGG freezes-in 
the ${\bf q}$-dependence of quasi-critical 
incommensurate 
correlated regions
as $T_g$ is crossed upon cooling.
We therefore use a two-step fitting procedure 
to determine $J_2$ and $J_3$.
First, we establish whether the model ${\cal H}$
allows for a ${\bf q}_{\rm ord}$ compatible with the experiment. 
Second, we examine whether the model
with a given $\{J_2,J_3\}$ set is  
able to reproduce 
the overall structure of correlations 
both in the paramagnetic and ``frozen-in critical'' 
regimes~\cite{Petrenko-PRL}. 

We find that ${\bf q}_{\rm ord}$ for $R_c=\infty$ 
belongs to the (hhl) reciprocal plane, with
its location within the 
plane highly dependent on 
$J_2$ and $J_3$. 
To determine the optimum $J_2$ and $J_3$, 
we generate a sequence of data points 
$\{\tilde{Q}\} \equiv |{\bf q}_{\rm ord}(J_2,\, J_3)+ {\bf {G}}|$, 
and then match these points 
to the experimental sequence of sharp peaks in powder 
GGG data~\cite{Petrenko-PRL}.
To do so, we introduce a penalty function,
$\mathcal{P}(J_2,\, J_3)$,
that provides a measure of mismatch 
between the  first three clearly discernible and strongest 
experimental peaks at 
$Q_1^*=0.64$ \AA$^{-1}$, 
$Q_2^*=0.85$ \AA$^{-1}$ 
and 
$Q_3^*=1.07$ \AA$^{-1}$ of Ref.~\onlinecite{Petrenko-PRL} 
(see 43mK experimental data in top panel of Fig.~\ref{figIq})
and the closest theoretically determined peaks:
$\mathcal{P}= 100 \times \max_{i=1,2,3}\Delta Q_i$, where
$\Delta Q_i=\min_{\tilde{Q} \in \{\tilde{Q}\}}|\tilde{Q}-Q_i^*|$
with the numerical factor 100 chosen to set the scale, ${\mathcal{P}}\equiv 1$, 
for a maximum mismatch of 0.01\AA$^{-1}$.
A map of $\mathcal{P}$
in the $J_2-J_3$ plane is shown in 
Fig.~\ref{figJ2J3map}.
A good match between theory and experiment is denoted in black.
The discrepancy between experiment and theory increases
as the 
gradation of gray varies from black to white.

\begin{figure}[t]
\begin{picture}(250,150)
\put(0,185){\includegraphics[width=70mm,angle=-90]{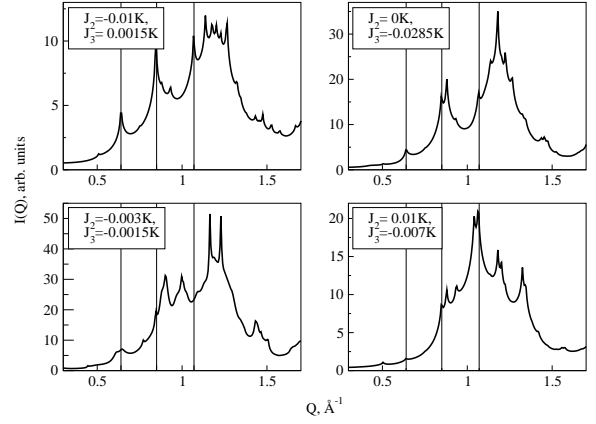}}
\end{picture}
\caption{
Scattering profile 
for regions of 
the $J_2-J_3$ plane where the positions $Q_i$ of the sharp peaks have 
been optimized; $\tau=1.6\times 10^{-6}$. 
Upper left and bottom right panels represent correspondingly located 
black domains 
in Fig.~\ref{figJ2J3map}.
The other two panels represent small domains in the central and
bottom-left part of Fig.~\ref{figJ2J3map}. The overall shape of $I(Q)$
varies  significantly, but all
the profiles
 display sharp peaks agreeing with
the experiment (shown by vertical lines).
}
\label{figshape}
\end{figure}
\begin{figure}[t]
\includegraphics[angle=-90,width=2.8 in]{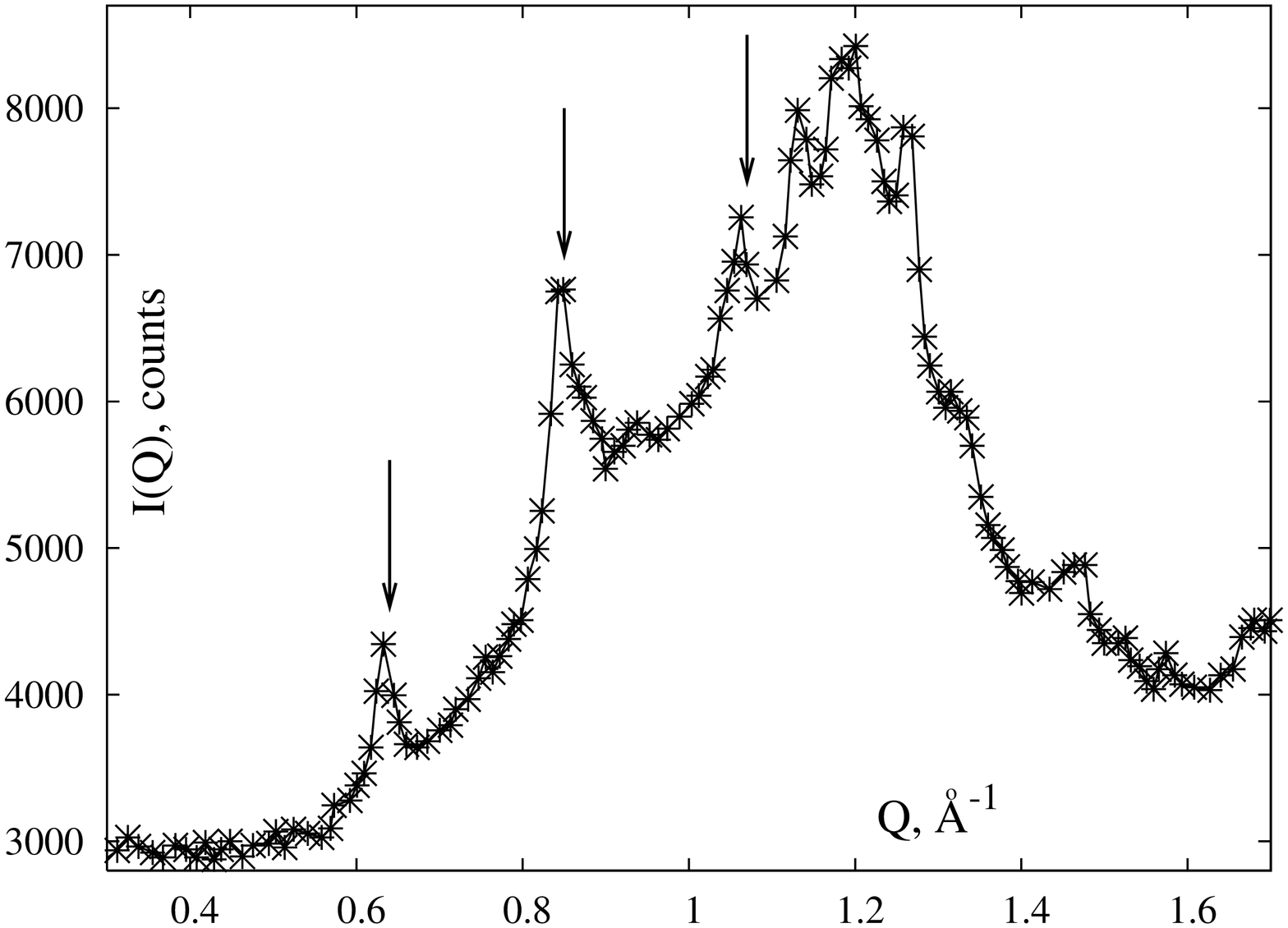}
\includegraphics[angle=-90,width=2.8 in]{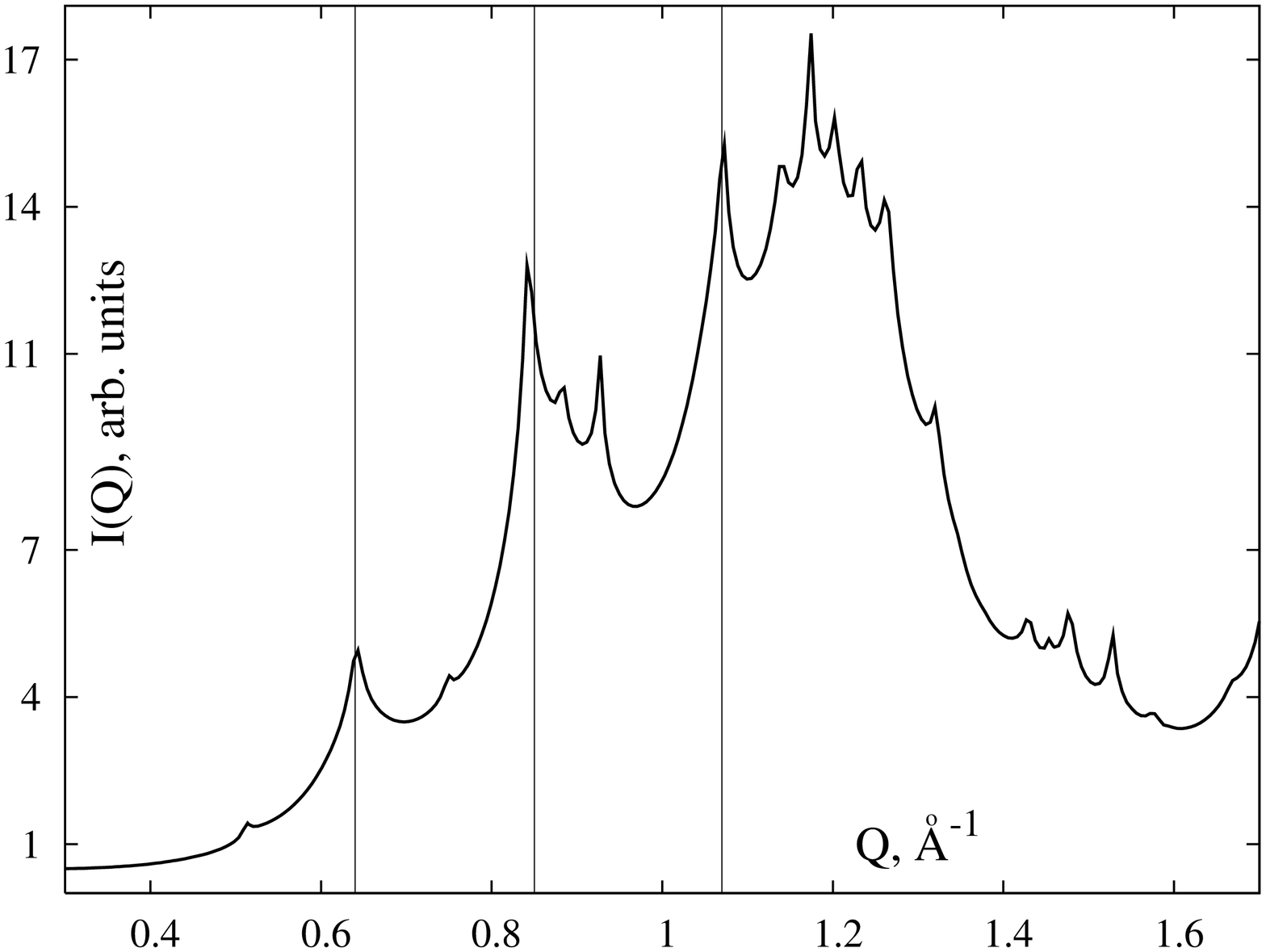}
\caption{
\label{figIq}
Positions of the first three strongest peaks of the experimental
scattering profile (pointed to by arrows in the upper panel) 
as well as the overall shape of scattering profile
are used as criteria for the optimization procedure in the $J_2-J_3$ plane. 
The resulting mean-field theory profile at 
dimensionless temperature $\tau=1.6\times 10^{-6}$ 
(lower panel) 
compares well with
the experimental one.
The positions of three experimental peaks are denoted by vertical lines.
The experimental scattering profile has been obtained by digitizing 
Fig.~2 of Ref.~\onlinecite{Petrenko-PRL}.
The theoretical profile is shown for $J_2=-0.005$K, $J_3=0.010$K,
denoted by a box in Fig.~\ref{figJ2J3map}.}
\end{figure}

Figure~\ref{figJ2J3map} shows that 
even small (at the $\approx 1\%$ level of $J_1$) changes in $J_2$ and $J_3$ 
show up as large changes in the ordering wave vector and, correspondingly, in
$\mathcal{P}$ too. There exist only relatively
small domains 
(shown in black)
that minimize 
$\mathcal{P}$.
It turns out, however, that
for each of these 
domains,
the overall shape of the MFT powder 
scattering profile changes noticeably 
even though 
${\bf {q}}_{\rm ord}$ remains almost 
constant at the value of 
$2\pi/a\, [0.29, 0.29, 0]$.
Similarly to the standard
experimental procedure to analyze powder neutron scattering data,
once ${\bf q}_{\rm ord}$ is determined,
one needs to solve for the magnetic structure.
The sensitivity
of the unit cell magnetic form factor to $\{J_2,J_3\}$
leads to a significant {\it qualitative} 
change in the diffuse scattering pattern near
criticality, hence offering a second optimization channel
for the determination of $\{J_2,J_3\}$.

Figure~\ref{figshape} shows the dependence of the theoretical $I(Q)$ on
$\{ J_2,J_3 \}$ within the black domains
of Fig.~\ref{figJ2J3map}. 
A comparison of such theoretical profiles with the experimental one 
reveals that {\it only} in the upper left black
domain are the theoretical 
and experimental profiles similar; a strong discrepancy in other domains
allows us to exclude them from further consideration.
Within the upper left domain, 
the match is visually better for rightmost values of 
$J_2$ with the restriction that its value 
cannot cross the domain boundary at $-0.004$K.

We support and further refine the above analysis 
by fitting the MFT scattering profiles to the experimental  
$I(Q)$ in the paramagnetic regime. 
For each value of $J_2$ and $J_3$ we consider a 
$\chi^2$-type penalty function
for the overall shape of $I(Q)$, which is adjusted by parameters responsible
for a uniform and linear in $|Q|$ background terms, as well as by the
overall scale factor. 
With $\tau$ as a free fitting parameter,
we do the fits in the interval $Q=0.26-1.57$\AA$^{-1}$
which allows for the complete shape 
determination of the first experimental broad 
diffusive scattering peak at 1.05\AA$^{-1}$
without a $|Q|^2$ background correction.
Most importantly,
an analysis of the (paramagnetic) data at 175~mK~\cite{Petrenko-PRL} 
allows us to confirm the outcome 
of the analysis of the critical data at 
43 mK and reject all but the aforementioned upper left domain
of Fig.~\ref{figJ2J3map}.
Within that domain 
the data impose conservative error bars on $J_2$: 
$-0.012~{\rm K}\leq J_2 \leq -0.004~{\rm K}$, 
with limits on $J_3$ set by that same domain,
i.e. $-0.003~{\rm K}\leq J_3 \leq 0.012~{\rm K}$, 
see Fig.~\ref{figJ2J3map}.

The MFT $I(Q)$
with the optimized $J_n$ values 
captures the features of
the experimental $I(Q)$ 
well (see bottom panel of Fig.~\ref{figIq}).
The optimization procedure to match the three strongest experimental
peaks leads to a theoretical sequence of peaks
that can be identified on the experimental profile as 
rather well-distinguished peaks and cusps.
The match (Fig.~\ref{figIq}) between 
MFT and experimental
$I(Q)$ 
suggests post-factum that the 
${\bf q}_{\rm ord}$ dependence of the
correlations 
can be described by MFT. 

To conclude, we have found that, as in 
spin ices~\cite{Diep-review,Gingras-CJP,Melko},
an ad-hoc $R_c$ cut-off  of
the dipolar interactions at less than a few
hundred nearest-neighbor distances leads to spurious 
long-range ordered phases.
Only a proper treatment of an infinite $R_c$ allows one
to get a handle on the very small exchange interactions
beyond nearest neighbors 
which dictate the incommensurate ordering wave vector 
${\bf q}_{\rm ord}$.
With a reasonably  well parameterized
Hamiltonian now in hand, further theoretical studies to explore
the origins of the complex phenomena displayed by GGG in zero
and nonzero magnetic field may now be possible.
Perhaps most importantly, we have identified the likely
reciprocal plane that contains 
${\bf q}_{\rm ord}$, with 
${\bf q}_{\rm ord}\approx$ $2\pi/a\, [0.29, 0.29,0]$.
With this prediction available, it may now be possible to 
perform single crystal neutron scattering measurements in
a reflection geometry to beat the $^{157}$Gd absorption
problem
in high quality single crystals with natural Gd abundance~\cite{Gaulin}, 
and allow for a quantitative investigation of 
the development of correlations in GGG at low temperatures. 
This may prove a fruitful endeavor to help shed 
light on the mysteries of GGG.


We acknowledge the contributions 
of M. Dion and B. den Hertog to the earlier stage of this project.
We thank A. Del Maestro, 
B. Gaulin, R. Moessner, O.~Petrenko, A. Ramirez and P. Schiffer 
for useful discussions. We thank O. Petrenko for prodiving
neutron data.
This work was funded by NSERC
of Canada, the Canada Research Chair Program (Tier I) (M.G),
Research Corporation (M.G., M.E.), the Canada foundation for Innovation,
the Ontario Innovation Trust, and the Canadian Institute for Advanced research.
M.G. acknowledges the U. of Canterbury (UC) for an Erskine Fellowship
and the hospitality of the Dept. of Physics and Astronomy at UC
where part of this work was completed.


\end{document}